# A framework of learning controller with Lyapunov-based constraint and application

Me Le, Chi Yanxun, Xu Dongfu*, Li Zhiwei, Zhang Yulong

**Abstract:** In this paper, we focus on the problem about direct way to design a stable controller for nonlinear system. A framework of learning controller with Lyapunov-based constraint is proposed, which is intended to transform designing and analyis of a controller to straightforward way to make controller by solving an optimization with the Lyapunov constraint, and which can be a novel way to design a global stability guaranteed controller directly. Firstly, an optimization problem subject to Lyapunov-based constraints is formulated, in which the tracking error is the objective function to minimize. Secondly, the controller combines with PID and feedforward is given in form of neural networks, Finally, the solution of controller method to the optimization problem is analyzed, in which we leverage some deep learning technologies to boost the capbility of solution. The results of two simulations of 2 order linear and nonlinear systems demonstrate that the method proposed has some high performance in speed of convergence, tracking error and smoothness and amplitude of control output. The results of simulation to the nonlinear system with disturbance, noise, uncertainty of parameters and the difference of reference output demostrate that our method has high performance in robustness and generalization.

# 一种基于 Lyapunov 约束的学习控制方法及应用

马乐，迟延迅，徐东甫*，李志伟，张玉龙
东北电力大学 自动化工程学院 机器人研究所

**摘要：**针对非线性系统的稳定控制器直接设计问题，提出一种基于 Lyapunov 稳定性条件的学习控制器设计方法框架，将传统的控制器设计与稳定性证明分析问题转化为以控制器为求解项，Lyapunov 稳定条件为约束的最优化问题，提供了一种直接求解全局稳定的最优控制器的新途径。首先建立了系统跟踪误差与 Lyapunov 稳定约束的最优化问题，接着给出了一类基于神经网络实现的 PID 结合前馈控制器设计形式，最后分析并设计了学习控制器求解方法，采用相关深度学习技术以提升求解能力。二阶线性与非线性系统仿真测试表明本文提出方法具有快速收敛、低误差、控制输出平滑且低幅值等特点；在加入扰动、噪声、参数不确定性和与学习阶段不同的测试期望输出条件下的测试结果表明本文方法具有对扰动与噪声具有良好的抑制能力同时学习控制器具有良好的泛化能力。

关键词：直接控制器设计；Lyapunov 约束；学习控制；深度学习技术

## 1 引言

一直以来控制器的系统稳定性分析在控制理论及应用中占有重要地位，稳定性是控制器实际工作的先决条件[1]。

目前控制理论研究通常做法是先设计针对特定模型系统的控制器，后分析其稳定性。然而在控制器设计过程中，或因稍过于考虑稳定性分析而忽略了实际的控制性能。同时普遍存在以下问题：1）稳定分析对系统与控制器参数依赖，当实际与分析存在参数差异时不能保证分析仍成立，或需重调参数以确保稳定；2）稳定性分析中通常忽略或简化虑噪声与扰动作用导致控制器在上述条件下的动态性能有待提升[2]；3）传统控制方法完成控制器设计后对系统的控制性能随之固定缺乏性能提升途径。因此产生下述动机：是否能够瞄准控制问题本身直接设计具性能提升能力与全局稳定性的控制器？

通常 Lyapunov 函数被用于控制器的稳定性分析，将其作为控制器设计条件是上述动机的积极思路[3]。现有研究中下述控制方法蕴含上述思想。首先 Backstepping 控制引入了“虚拟控制”针对严格反馈系统逐级建立 Lyapunov 函数并设计相应虚拟控制量实现最终控制器设计[4]。然而该方法局限于系统模型形式，并常伴随一定的稳态误差，同时当期望输出不确定时其导数难以精确获得，为此[5]等研究引入最优条件力图改善控制性能，但该方法仅适用于凸优化问题，往往该条件难以保证。

另一种研究方法是寻找系统所谓的“控制 Lyapunov 函数(Control Lyapunov Function, CLF)”[6]并利用 Sontag 公式直接构造稳定控制器[7]。同样该方法存在一定的局限，首先方法仅适用于仿射系统[8]，其次寻找系统的控制 Lyapunov 函数往往与稳定性分析同样困难[9]。因此该方法虽提供了一个直接的控制器设计途径，但未从根本上简化控制设计。

另外注意到控制问题根本问题之一是跟踪误差最小化，因此引入最优化指标也是实现上述动机的必要方法[4, 10]。模型预测控制(Model Predictive Control, MPC)一类基于求解最优化的控制方法，而然因传统 MPC 的预测范围有限，通过求解有限时段内最优化不能自动的确保闭环稳定性[11]。为此[12]等研究引入了 Lyapunov 稳定性条件。其中[13]在传统 MPC 优化环节加入了 CLF 约束项以实现稳定性约束，然而该方法需建立在某个已具备 Lyapunov 稳定控制器基础上，因此未能解决上述动机。

综上本文提出了一种以 Lyapunov 稳定性为约束条件的学习控制器设计框架，将跟踪误差范数与系统 Lyapunov 函数导数约束联立构成最优化问题，并依此求解最优学习控制器。本文给出了一个由神经网络实现的 PID、前馈控制器，采用一系列深度学习技术以提升对优化问题的求解能力。本文贡献如下：

1）本文方法框架通过引入 Lyapunov 稳定条件约束，提供一种直接稳定控制器的设计的途径；

2）方法中采用的控制输出限幅技术使得到的控制器幅值自然限于给定界限内，规避了控制输出饱和问题；

3）本文采用的深度学习相关技术提升了优化问题的，同时学习过程为单点优化有利于方法的实际应用；

4）经学习后的控制器具有良好的泛化能力，能脱离学习阶段独立运行。

## 2 问题描述

考虑控制系统如：

$$\begin{cases}\dot{\mathbf{x}} = \mathbf{F}(\mathbf{x},\mathbf{u}) \\ \mathbf{y} = \mathbf{H}(\mathbf{x})\end{cases} \qquad (1),$$

选择 Lyapunov 函数 $V(\mathbf{x})$,若设计的控制器 $\mathbf{u}(\mathbf{x}\,|\,\boldsymbol{\theta}^{\mathbf{u}})$ 使 $\dot{V} < 0$,则 $\mathbf{u}(\mathbf{x}\,|\,\boldsymbol{\theta}^{\mathbf{u}})$ 对控制系统全

局稳定。因此建立不等式约束最优化问题：

$$\begin{cases}\min \left\|\mathbf{y}^{\mathrm{d}}-\mathbf{y}\right\| \\ \text{s.t. } \dot{V}<0\end{cases} \tag{2}$$

其中$\min \left\|\mathbf{y}^{\mathrm{d}}-\mathbf{y}\right\|$意在使控制系统输出的跟踪误差最小，$\dot{V}<0$作为控制过程的全局稳定约束条件。若(2)式问题存在最优解$\mathbf{u}^{*}$，则$\mathbf{u}^{*}$为 Lyapunov 全局稳定条件下的最优控制器。

取跟踪误差内积为 Lyapunov 函数

$$V(\mathbf{x})=\left(\mathbf{y}^{\mathrm{d}}-\mathbf{y}\right)^{\mathrm{T}}\left(\mathbf{y}^{\mathrm{d}}-\mathbf{y}\right) \tag{3}$$

考虑离散化并代入$\dot{\mathbf{x}}=\mathbf{F}(\mathbf{x},\mathbf{u})$得：

$$\begin{cases}\min \left\|\mathbf{y}_{k}^{\mathrm{d}}-\mathbf{y}_{k}\right\| \\ \text{s.t.} \left(\dfrac{\partial \mathbf{V}}{\partial \mathbf{x}}\right)^{\mathrm{T}} \mathbf{F}\left(\mathbf{x}_{k}, \mathbf{u}(\mathbf{x}_{k} \mid \boldsymbol{\theta}^{\mathrm{u}})\right)<0 \\ \mathbf{y}_{\mathrm{k}}=\mathbf{H}\left(\mathbf{x}_{k}+\mathbf{F}\left(\mathbf{x}_{k\text{-}1}, \mathbf{u}(\mathbf{x}_{k\text{-}1} \mid \boldsymbol{\theta}^{\mathrm{u}})\right) \Delta t\right)\end{cases} \tag{4}$$

其中$\Delta t$为采用时间。

(2)(4)式将设计控制器并证其李雅普诺夫稳定问题转为 Lyapunov 稳定性为约束条件的优化问题，进而给出了一种直接求解全局稳定最优控制器的新途径。

**3 学习控制器设计**

本文重点在于提出的 Lyapunov 稳定约束的学习控制器设计框架，对于该框架可以设计不同的控制器结构。本文仅对一类神经网络控制器加以分析。

本文设计的学习控制器如下：

$$\mathbf{u}_{\mathrm{L}}=\mathbf{u}_{\mathrm{p}}+\mathbf{u}_{\mathrm{f}} \tag{5}$$

其中$\mathbf{u}_{\mathrm{p}}$与$\mathbf{u}_{\mathrm{f}}$分别为 PID 与前馈控制器。区别于传统方法，上述三种控制项均被采用神经网络实现，记为$\mathbf{u}_{\mathrm{p}}(\mathbf{x}_{\mathrm{p}} \mid \boldsymbol{\theta}^{\mathrm{p}})$与$\mathbf{u}_{\mathrm{f}}(\mathbf{x}_{\mathrm{f}} \mid \boldsymbol{\theta}^{\mathrm{f}})$,其中$\boldsymbol{\theta}^{*}$为待学习的神经网络参数，因此控制器$\mathbf{u}_{\mathrm{L}}$的待学习参数为$\boldsymbol{\theta}^{\mathrm{L}}:\left\langle\boldsymbol{\theta}^{\mathrm{p}}, \boldsymbol{\theta}^{\mathrm{f}}\right\rangle$。

本文采用三层前向网络构建学习控制器，其中隐层激活函数选为 sigmoid 与线性整流函数(Rectified Linear Unit, ReLU):

$$\begin{cases}f_{\text{sigmoid}}(x)=\dfrac{1}{1+e^{-x}} \\ f_{\text{ReLU}}(x)=\max(0, x)\end{cases} \tag{6}$$

$\mathbf{u}_{\mathrm{p}}$输入量为系统输出的误差、积分项与微分项，记为$\mathbf{x}_{\mathrm{p}}=\left[x_{\mathrm{e}}, x_{\mathrm{s}}, x_{\mathrm{d}}\right]^{\mathrm{T}}$；$\mathbf{u}_{\mathrm{f}}$输入由系统期望输入序列构成，记为$\mathbf{x}_{\mathrm{f}}=\left[y_{k}^{\mathrm{d}}, y_{k-1}^{\mathrm{d}}, \ldots\ldots, y_{k-N_{\mathrm{f}}+1}^{\mathrm{d}}\right]^{\mathrm{T}}$其中$N_{\mathrm{f}}$为输入维数。

仿真与实际系统中控制输入均应有界因此本文利用 tanh 函数对设计的神经网络控制器做限幅处理：

$$\begin{cases}\tilde{\mathbf{u}}_{\mathrm{L}}=\Delta V\cdot f_{\tanh}\left(\dfrac{\mathbf{u}_{\mathrm{L}}}{\Delta V}\right)+\dfrac{V_{\max}+V_{\min}}{2}\\ \Delta V=V_{\max}-V_{\min}\end{cases}\tag{7}$$

其中 $V_{\max}$ $V_{\min}$ 分别为系统控制输入上下限。该限幅方法优点在于相对 $\max\left(V_{\min},\min\left(V_{\max},x\right)\right)$，(7)式确保了对控制输出的平滑限制，有利于后续基于梯度的优化求解。

**4 学习控制器求解**

在确定学习控制器结构后，下一个关键问题是如何利用控制系统信息优化控制器参数 $\boldsymbol{\theta}^{\mathrm{L}}$ 。

(3)式为一类带不等式约束的最优化问题，因此需将其转化为无约束优化问题。虽然 Lagrange 乘子法是一类通常采用的优化方法，但对于不等式约束问题需满足 KKT 条件[14]，然而(3)式不确保满足该条件。为此本文采用一种改进的惩罚函数法对(3)式非约束化。

建立惩罚函数 $\rho(x\,|\,\boldsymbol{\beta}_{\rho})$,其中 $\boldsymbol{\beta}_{\rho}$ 为函数参数,则不等式约束问题(3)可转化为无约束优化问题，目标函数为：

$$\begin{aligned}L(\boldsymbol{\theta}^{\mathrm{L}}\,|\,\cdot)&=\left\|\mathbf{y}_k^{\mathrm{d}}-\mathbf{y}_k\right\|\\&+\lambda\cdot\rho\left(\left(\frac{\partial\mathbf{V}}{\partial\mathbf{x}}\right)^{\mathrm{T}}\mathbf{F}\left(\mathbf{x}_k,\mathbf{u}_{\mathrm{L}}(\mathbf{x}_k\,|\,\boldsymbol{\theta}^{\mathrm{L}}\right)\,|\,\boldsymbol{\beta}_{\rho}\right)\end{aligned}\tag{8}$$

其中 $\lambda\,(>0\,)$为权衡系数。

惩罚函数 $\rho(x\,|\,\boldsymbol{\beta}_{\rho})$ 作用是当约束条件满足时函数不对优化问题产生影响，而当不满足要求时产生一个惩罚项，迫使待优化参数趋向可行域。因此目标函数 $L$ 最优解必同时满足跟踪误差极小与不等式约束两个条件。

为此本文将 $\rho(x\,|\,\boldsymbol{\beta}_{\rho})$ 设计为：

$$\begin{cases}\rho(x\,|\,[\beta_{\mathrm{B}},\alpha_{\rho}])=\left(f_{\mathrm{softplus}}(x+\beta_{\mathrm{B}})\right)^{\alpha_{\rho}}\\ f_{\mathrm{softplus}}(x)=\ln\left(1+e^{x}\right)\end{cases}\tag{9}$$

由(9)可看出当被约束项<0 时惩罚项趋近 0 对优化不产生重要影响，但当被约束项>0 时惩罚项近似呈幂函数增长，进而迫使优化器将待优化参数拉近约束条件域。(9)式中 $\beta_{\mathrm{B}}>0$ 为边界裕量为使约束项稍远离临界值 0。

至此，可采用基于梯度的优化算法（如 SGD、Adagrad、Adam 等）求解神经网络学习控制器参数。为保证在小误差状态下迭代性能，本文选择跟踪误差范数为 SmoothL1 范数[15]：（注：下式仅学习控制阶段求优过程使用，后续仿真结果均采用传统的跟踪误差指标评价）

$$\|x\|_{\mathrm{smoothL1}} = \begin{cases} \frac{1}{2}x^2, |x| < 1 \\ |x| - \frac{1}{2}, \text{otherwise} \end{cases} \tag{10}$$

**4 仿真分析**

本文采用 Python 与 Pytorch 框架进行仿真。仿真测试不同系统模型的控制效果以验证本文提出方法的有效性。

算例 1：二阶线性系统如下：

$$\begin{cases} \dot{x}_1 = x_2 \\ \dot{x}_2 = -Ax_2 - Bx_1 + Cu \\ y = x_1 \end{cases} \tag{11}$$

其中 $A=9$， $B=6$， $C=6$，期望轨迹设为 $y_t^{\mathrm{d}} = 2\sin(3\mathrm{t})$ ,采用周期为 $\Delta t = 0.01\mathrm{s}$ 。控制器相关参数设置为：$\lambda = 1$，$\boldsymbol{\beta}_{\rho} = [0.1, 1]^{\mathrm{T}}$，$V_{\max}$ $V_{\min}$ 分别为±50，采用 Adagrad 算法迭代训练，学习律 $\eta = 0.36$ 。

在初始状态 $[x_1, x_2] = [0, 0]$ 条件下仿真控制 30s，图 1 为本文方法对(11)式系统仿真控制效果。从图(a)中看出系统输出在<2.5s 时间内即达到稳定，并在<7.8s 时间内跟踪误差降至低误差范围内，同时能在初始动态保持较低超调量。从图(b)看出整个控制过程中 $u$ 始终保持在低幅值范围内，并且时序平滑。从图(c)看出跟踪误差范围在 10s 内降至 0.05 以下。综上本文方法对算例 1 的学习控制中具备超调小、收敛快速、跟踪精度高等特征。

图 2 为控制器完成学习过程后的控制过程结果，从图(a)看出控制器经过学习后在<1s 时间内即达到稳定，且动态性能较学习阶段明显提升，整个控制过程保持很低的误差，图(b)显示学习后控制输入依然保持低幅值平滑，图(c)显示控制过程中跟踪误差低于 0.01。综上经学习后控制器的性能相对完成学习之前性能有提高，并保持了很高的输入与跟踪误差性能。

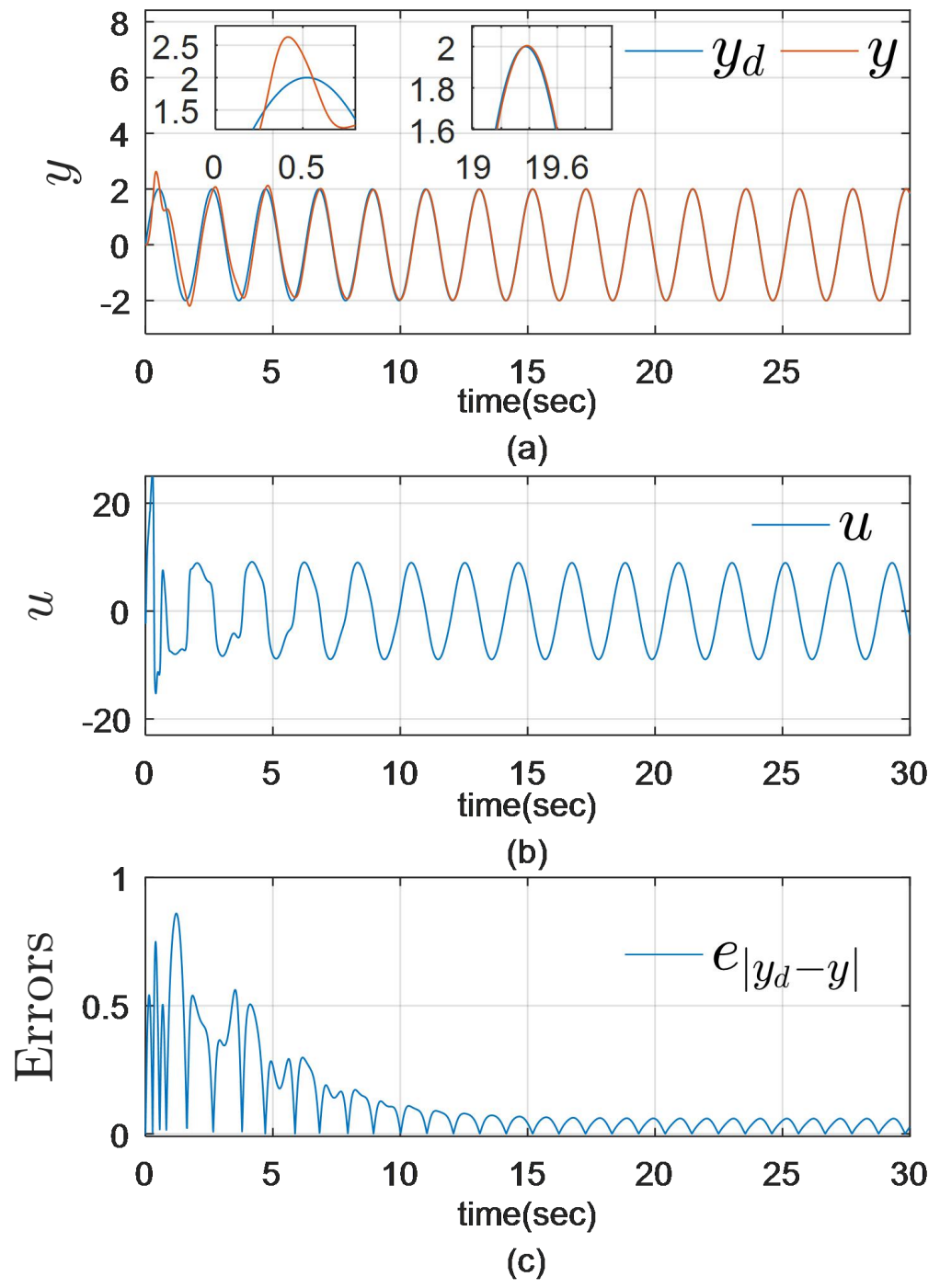

图 1 算例 1 学习控制结果

Fig1 Learing controlling performance of exmaple 1

算例 1 仿真实验证明了提出方法对于一类二阶线性系统控制具有良好控制特征。

算例 2：二阶线非性系统如下：

$$\begin{cases} \dot{x}_1 = x_2 \\ \dot{x}_2 = -A\cos(x_2) - B\sin(x_1) + Cu \end{cases} \tag{12}$$

本算例相关参数沿用算例 1，图 3 为本文方法对(12)式系统仿真控制效果。从图(a)中看出系统输出在<2.5s 时间内即达到稳定，并在<5s 时间内跟踪误差降至低误差范围内，由于系统非线性在初始学习阶段的超调大于算例 1 结果，但在较低范围内。从图(b)看出控制输出保持在低幅值，但由于系统非线性在学习控制器过程中输出较算例 1 有明显的震荡，分析原因为系统与 $y^{\mathrm{d}}$ 信号作用使控制信号频率增加，并受 Lyapunov 约束项的压迫导致学习过程中的震荡。图(c)表明对于该类非线性系统在较短的学习控制时间内仍能使跟踪误差收敛至很低范围。

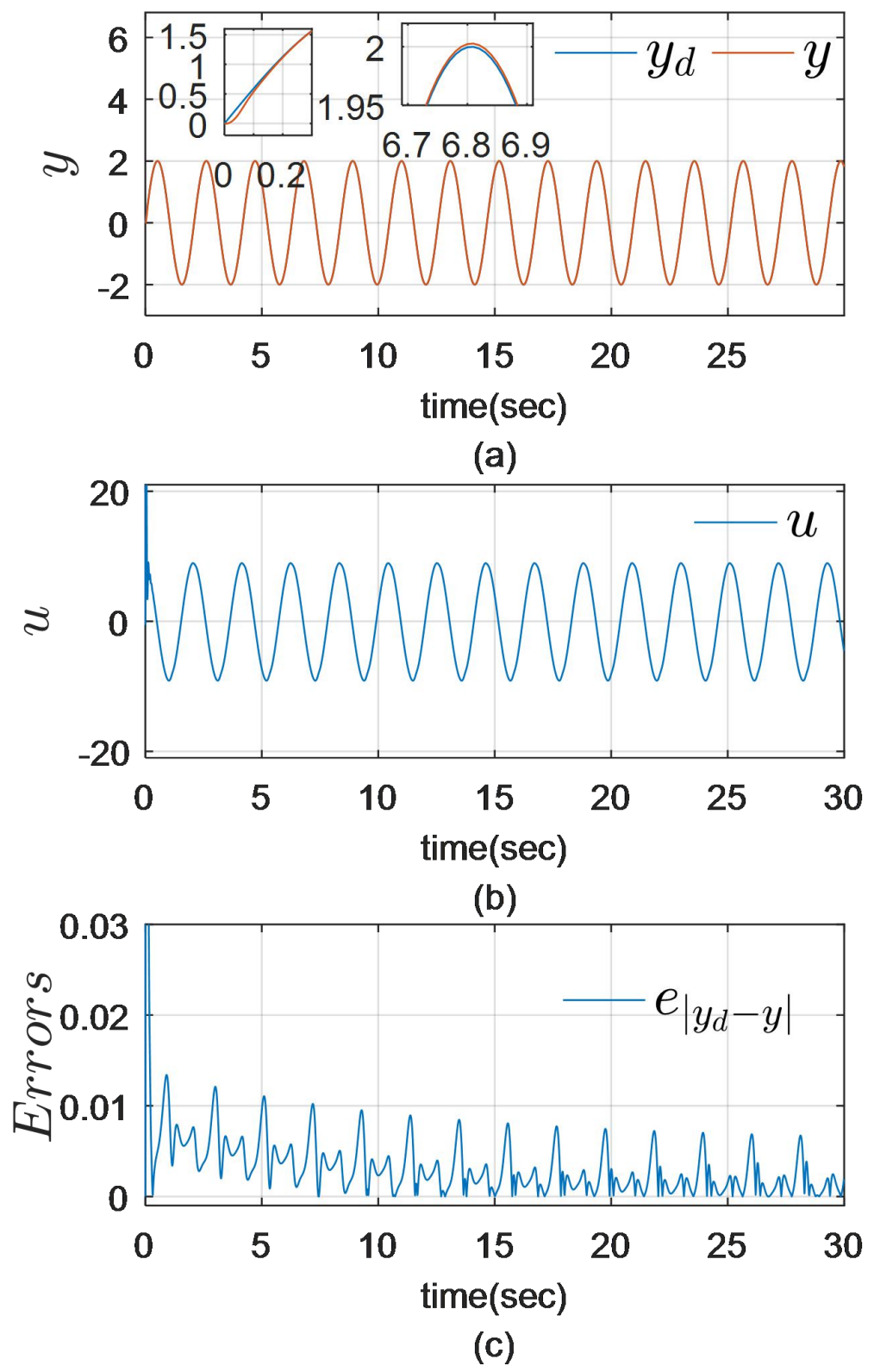

图 2 学习完成后控制结果

Fig 2 Test performance of exmaple 1 after learning

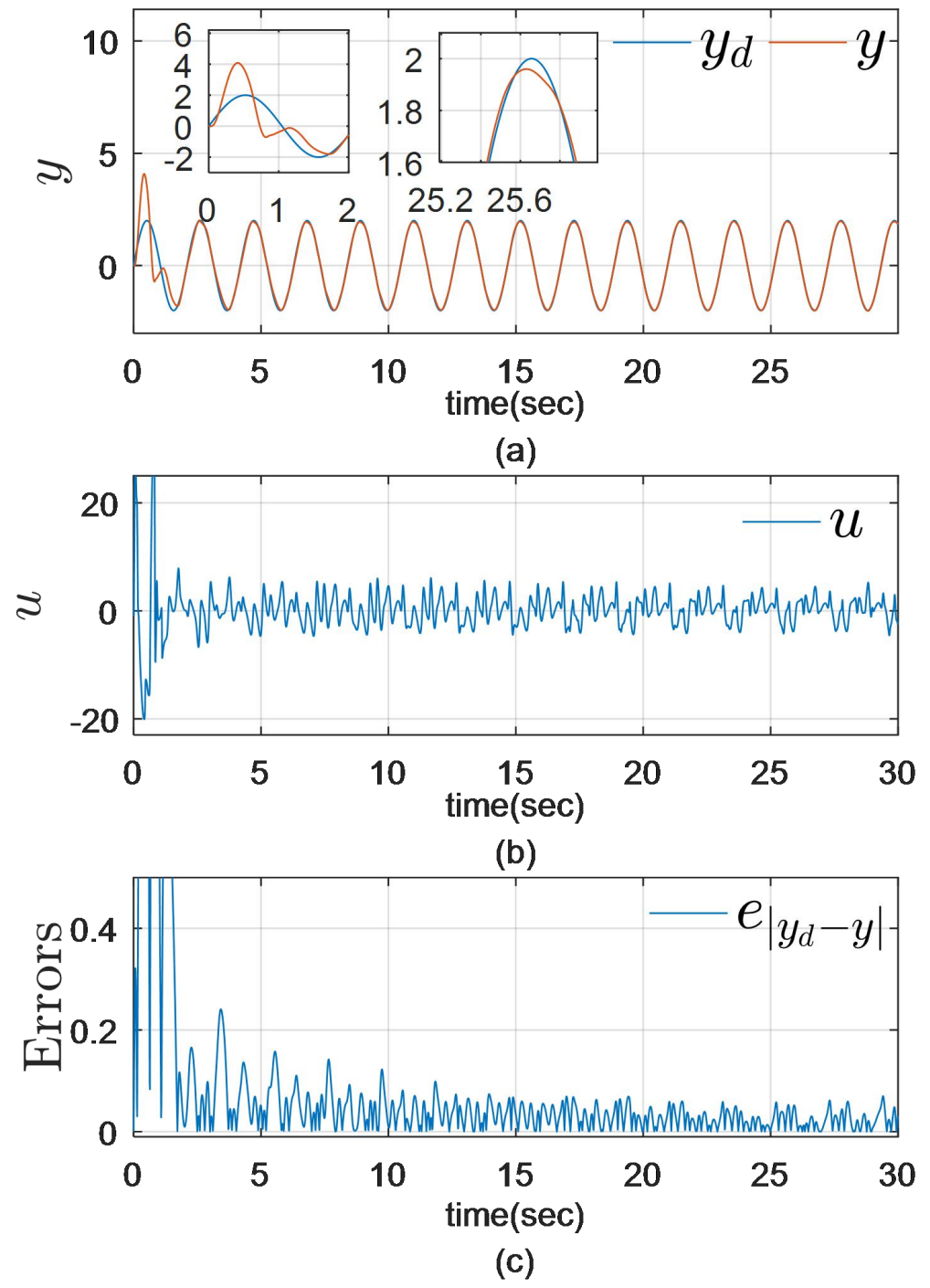

图 3 算例 2 学习控制结果

Fig 3 Learing controlling performance of exmaple 2

图 4 为控制器完成学习过程后的控制过程结果，从图(a)看出学习后的跟随控制效果在初始阶段即优于学习控制过程，并且整体跟踪效果良好，如图(b)所示学习后的控制输出震

荡明显小于学习过程，但由于非线性系统的频率叠加控制输出震荡高于算例 1；图(c)表明虽然跟踪误差整体高于算例 1 但仍保持低误差水平。综述本文方法对于该类非线性系统具有良好的学习能力与控制性能。

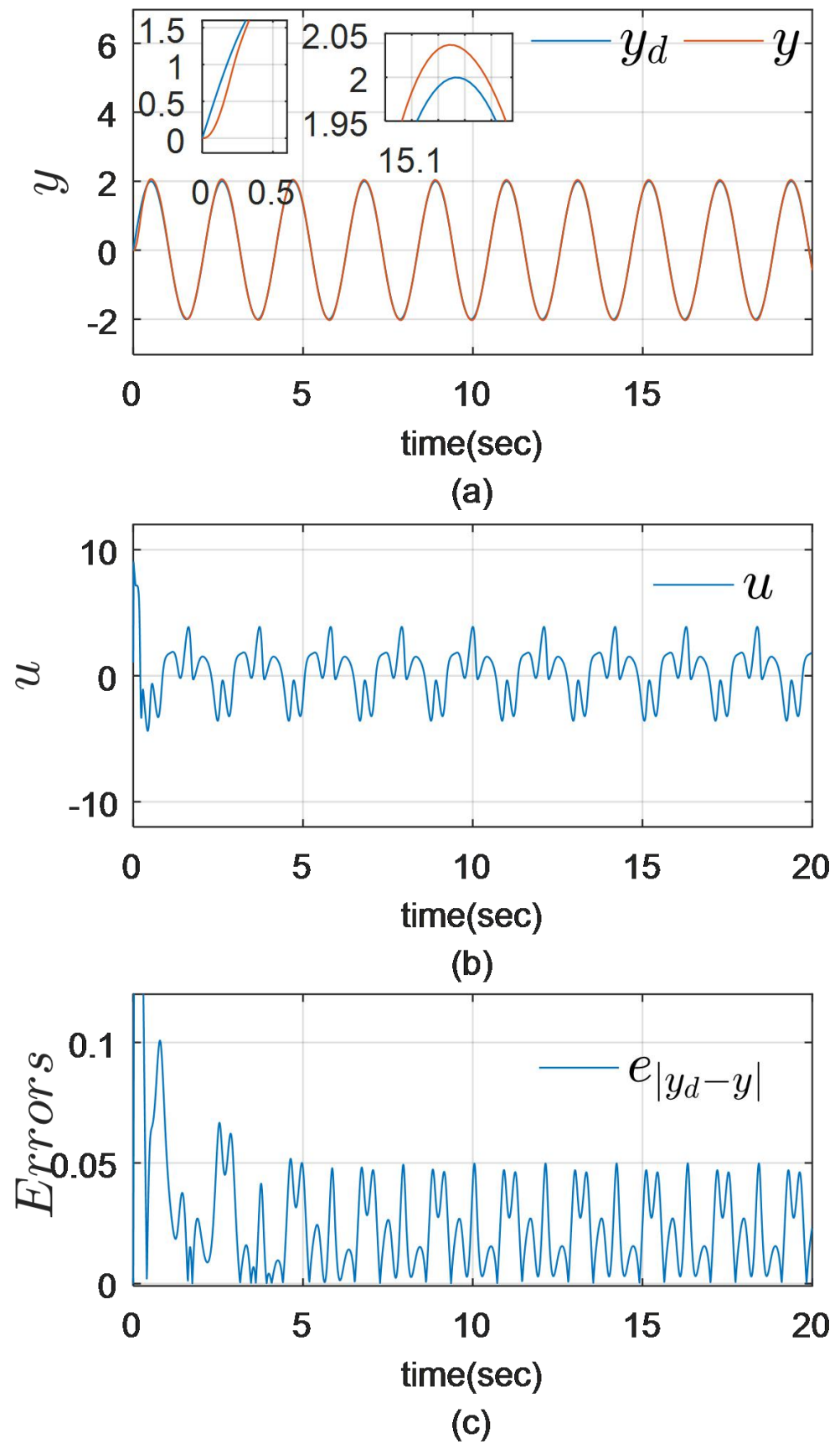

图 4 学习完成后控制结果

Fig4 Test performance of exmaple 2 after learning

算例 3：在算例 2 基础上对模型加入随机扰动 $d_{\mathrm{m}}$ 与噪声 $n_{\mathrm{m}}$ 以测试本文方法的抗噪、扰能力，模型形式如下：

$$\begin{cases} \dot{x}_1 = x_2 \\ \dot{x}_2 = -A\cos x_2 - B\sin x_1 + Cu + d_{\mathrm{m}} \\ y_{\mathrm{h}} = x_1 + n_{\mathrm{m}} \end{cases} \tag{13}$$

其中 $d_{\mathrm{m}}$ 与 $n_{\mathrm{m}}$ 分别为 $\pm 60$ 与 $\pm 0.2$ 的随机量。

同时测试阶段采用与学习阶段不同的模型与期望输出参数以测试本文方法对模型参数不确定的鲁棒性与学习算法的泛化能力，其中测试采用模型参数为 $A_{\mathrm{test}} = 8$， $B_{\mathrm{test}} = 6$， $C_{\mathrm{test}} = 3$，期望输出设为 ${}^{\mathrm{test}}y_t^{\mathrm{d}} = 1.5\sin(6t)$。图 5 为测试过程中噪声与扰动信号时序。

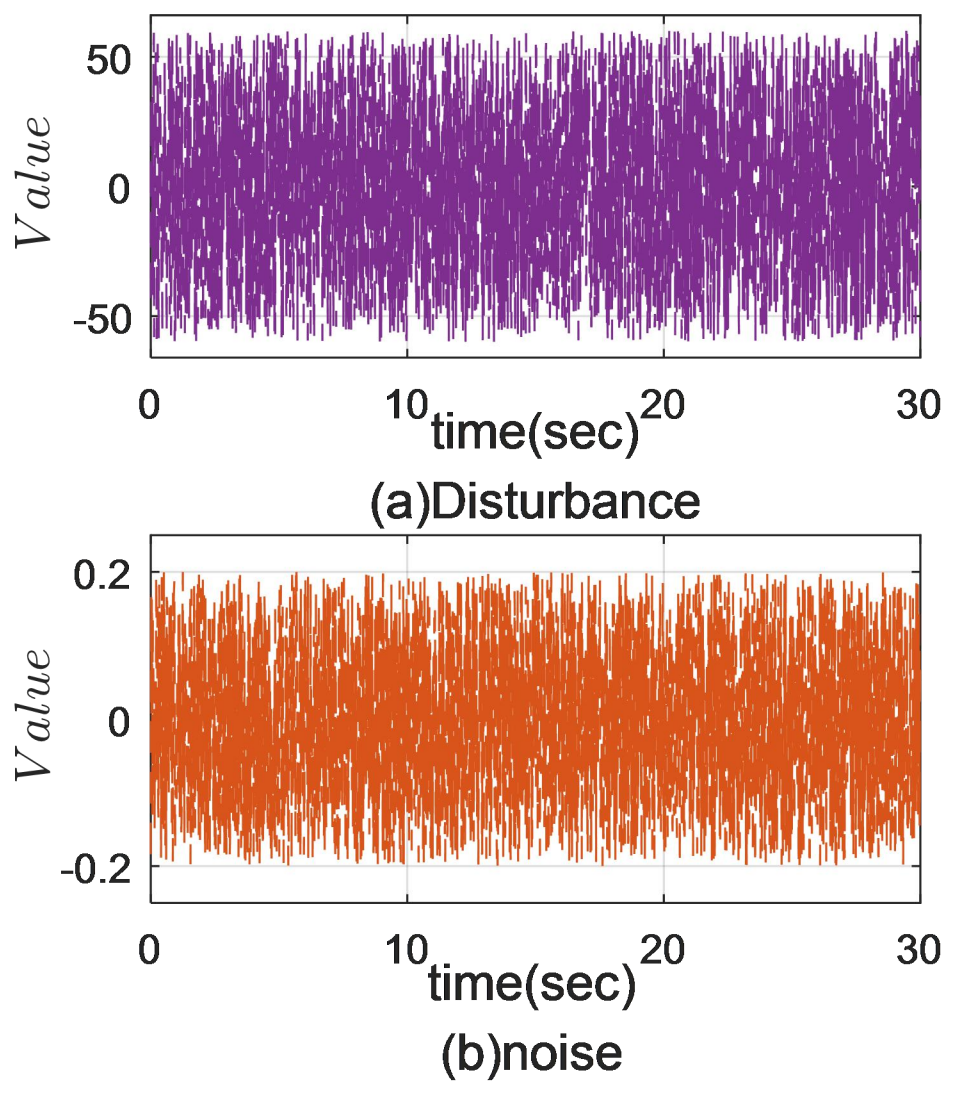

图 5  算例 3 控制过程中的噪声与扰动

Fig 5 Disturbance and noise in test of example 3

图 6 为控制器经学习后在测试条件下的控制效果。从图中看出噪声与扰动条件下学习后的控制器在不同模型与期望输出参数条件下能够较好的完成跟踪控制，具有较好的抗扰能力(c)图中的跟踪误差幅值低于噪声幅值；由于扰动与噪声存在加之期望输出频率高于学习阶段因此测试过程中的控制输出幅值与频率整体高于算例 2 结果。算例 3 结果表明，对于该类非线性系统本文方法具有良好的抗噪、扰能力、对参数不确定性的鲁棒性和学习算法的泛化能力。

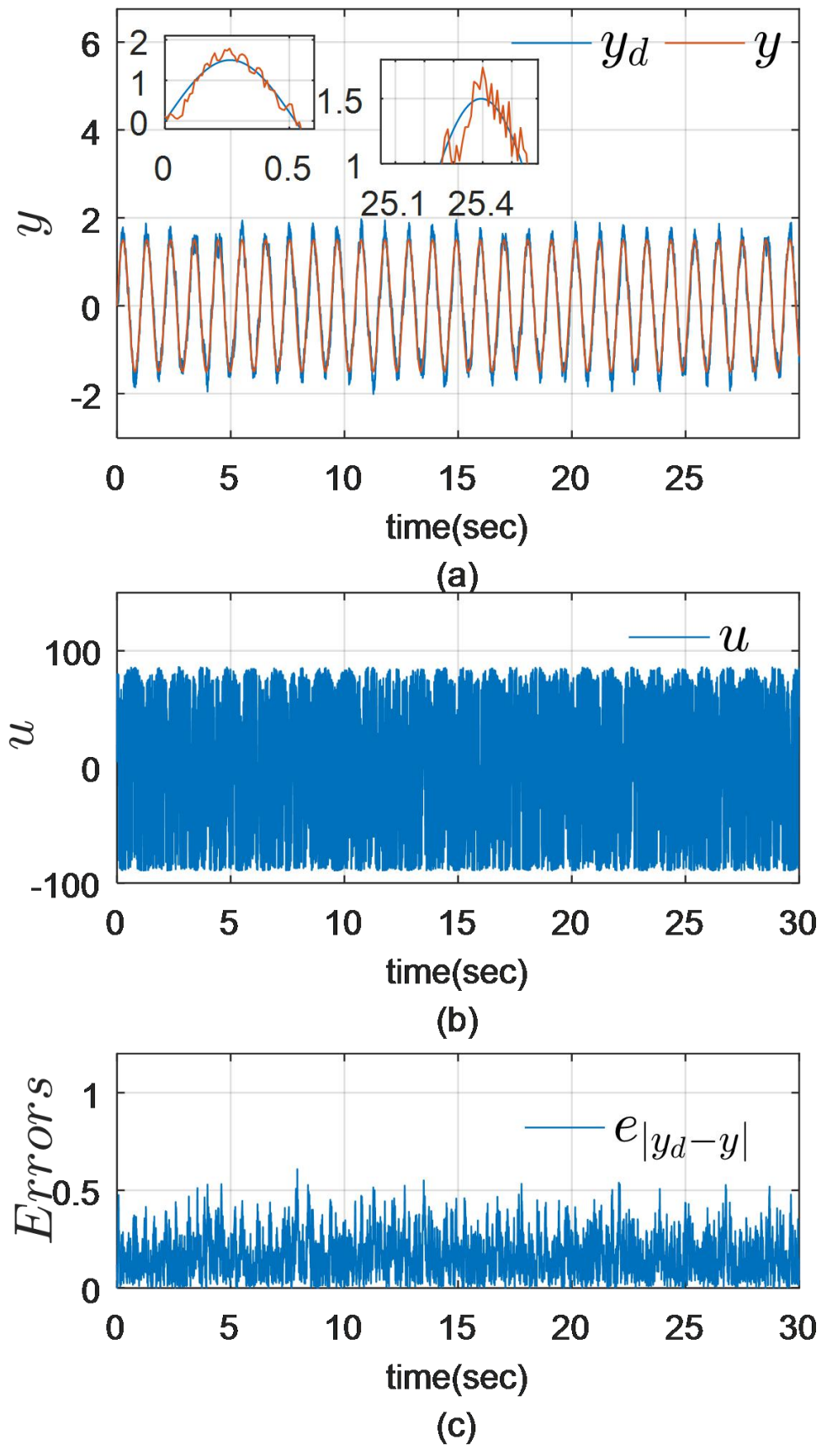

图 6  算例 3 测试结果

Fig 6 Test performance of exmaple 3 after learning

通过上述仿真测试验证了本文方法在不同类型系统控制上的有效性，并显式了在噪声、扰动、参数不确定性与学习泛化性等问题的处理能力。

## 6 结论

通过上述分析与仿真测试得出以下结论：1）本文提出的基于 Lyapunov 稳定性约束的学习控制器设计方法框架对于测试算例类型的二阶线性、非线性系统具有良好的控制器学习能力，能够直接的学习到控制品质良好并具全局稳定性的神经网络控制器；2）采用的控制输出限幅技术能够保证控制器输出在限定范围内进而规避了输出饱和问题同时满足实际控制需要；3）采用的深度学习优化技术与惩罚函数正则技术展现了良好的优化性能进而提升的控制器求解能力。上述结论表明了提出方法的有效性与先进性。

本文研究内容为引言动机的基础工作，如下问题值得分析与后续研究：1) 本文采用了基于 PID 参数与前馈为输入的控制器形式，因此学习阶段可看成对该种控制器的参数寻优。同时控制器形式也是决定性能的关键因素，后续研究将分析其他控制器结构；2) 仿真测试中用到了系统模型结构与反馈状态信息，后续研究将利用辨识技术对方法做无模型化与输出反馈扩展；3)最后，后续研究将扩大仿真模型类型并将方法应用于实际系统，同时对方法适用范围与极限做理论分析。

# 参考文献

[1] Mattila J, Koivumäki J, Caldwell D G, et al. A Survey on Control of Hydraulic Robotic Manipulators with Projection to Future Trends[J]. IEEE/ASME Transactions on Mechatronics. 2017, 22(2): 669-680.

[2] Wang D, He H, Liu D. Adaptive Critic Nonlinear Robust Control: A Survey[J]. IEEE Transactions on Cybernetics. 2017, 47(10): 3429-3451.

[3] Homer T, Mhaskar P. Constrained control Lyapunov function-based control of nonlinear systems[J]. Systems & Control Letters. 2017, 110: 55-61.

[4] Wen G, Ge S S, Tu F. Optimized Backstepping for Tracking Control of Strict-Feedback Systems[J]. IEEE Transactions on Neural Networks & Learning Systems. 2018, 29(8): 1-13.

[5] Ascencio P, Astolfi A, Parisini T. Backstepping PDE Design: A Convex Optimization Approach[J]. IEEE Transactions on Automatic Control. 2017, PP(99): 1.

[6] Furqon R, Chen Y, Tanaka M, et al. An SOS-Based Control Lyapunov Function Design for Polynomial Fuzzy Control of Nonlinear Systems[J]. IEEE Transactions On Fuzzy Systems. 2017, 25(4): 775-787.

[7] Sontag E D. A ‘universal’ construction of Artstein's theorem on nonlinear stabilization[J]. Systems & Control Letters. 1989, 13(2): 117-123.

[8] Guerrero-Castellanos J F, Rifaï H, Arnez-Paniagua V, et al. Robust Active Disturbance Rejection Control via Control Lyapunov[J]. Control Engineering Practice. 2018, 80(2018): 49-60.

[9] 陈奕梅, 韩正之. 基于控制Lyapunov函数的鲁棒自适应控制器设计[J]. 系统工程与电子技术. 2006(03): 435-438.

[10] 惠宇, 池荣虎. 基于迭代扩张状态观测器的数据驱动最优迭代学习控制[J]. 控制理论与应用. 2018, 35(11): 1672-1679.

[11] Mayne D Q. Model predictive control: Recent developments and future promise[J]. Automatica.

2014, 50(12): 2967-2986.
[12] Chao S, Yang S, Buckham B. Trajectory Tracking Control of an Autonomous Underwater Vehicle using Lyapunov-based Model Predictive Control[J]. IEEE Transactions on Industrial Electronics. 2018, 65(7): 5796-5805.
[13] Pena D M D L, Christofides P D. Lyapunov-Based Model Predictive Control of Nonlinear Systems Subject to Data Losses[J]. IEEE Transactions on Automatic Control. 2008, 53(9): 2076-2089.
[14] Tang W, Daoutidis P. A bilevel programming approach to the convergence analysis of control-Lyapunov functions[J]. IEEE Transactions on Automatic Control. 2019, PP(99): 1.
[15] Liu W, Anguelov D, Erhan D, et al. SSD: Single Shot MultiBox Detector[C]. 2016.